\documentstyle[11pt,newpasp,twoside,epsf]{article}
\def\edcomment#1{\iffalse\marginpar{\raggedright\sl#1\/}\else\relax\fi}
\reversemarginpar

\begin{document}
\title{Ultra-Short Period Double-Degenerate Binaries}
\author{Mark Cropper, Gavin Ramsay, Kinwah Wu}
\affil{Mullard Space Science Laboratory, University College London, 
Holmbury St Mary, Dorking, Surrey RH5 6NT, United Kingdom}
\author{Pasi Hakala}
\affil{Observatory, P.O. Box 14, FIN-00014 University of Helsinki, Finland}

\begin{abstract}
We review the current observational status of the {\it ROSAT} sources
RX~J1914.4+2456 and RX~J0806.3+1527, and the evidence that these are
ultra-short period ($<10$ min) binary systems. We argue that an Intermediate
Polar interpretation can be ruled out, that they are indeed compact binaries
with a degenerate secondary, and that the period seen in the X-ray and
optical is the orbital period. A white dwarf primary is preferred, but a
neutron star cannot be excluded. We examine the capability of the three current
double-degenerate models (Polar, Direct Accretor and Electric Star) to
account for the observational characteristics of these systems. All models
have difficulties with some aspects of the observations, but none can be
excluded with confidence at present. The Electric Star model provides the best
description, but the lifetime of this phase requires further
investigation. These ultra-short period binaries will be strong gravitational
wave emitters in the {\it LISA} bandpass, and because of their known source
properties will be important early targets for gravitational wave studies.
\end{abstract}

\section{Introduction}

Binary systems consisting of interacting white dwarf pairs (double degenerate)
have been known for some time. Smak (1967) was the first to suggest that AM CVn
was such a system, and although many questioned this interpretation, it is now
generally accepted. There are four AM CVn binaries in Warner's (1995) review,
and approximately ten such systems are now known. They have been considered to
be the double-degenerate analogues of Cataclysmic Variables, but with a
degenerate rather than main sequence-type secondary, accreting by Roche Lobe
overflow through a disk.

In the first Cape Workshop, Motch et al (1995) reported the discovery of a soft
X-ray source, RX~J1914.4+2456 (now V407 Vul), from the {\it ROSAT} Galactic
Plane Survey. They reported this as a new `soft' Intermediate Polar. The X-ray
emission was to be modulated with an amplitude of 100\% at a period of
568s. Motch et al (1996) searched for the optical counterpart, but were unable
to identify any unusual object.

In followup observations of the `soft' IPs, Cropper et al (1998) were exercised
by the shape of the X-ray modulation in this particular object, and by the fact
that for half of the modulation phase there was no detectable X-ray
emission. They noted that this was unlike the behaviour of all other IPs and
proposed instead that the period was the orbital period of a rotationally
sychronised white dwarf pair with a magnetic primary -- the double degenerate
analog of Polars. The implication of this was that at 568s (9.5 min) this was
the shortest period binary system known.

Since then the optical counterpart has been identified (Ramsay et al 2000) and
significant progress has been made in elucidating the nature of the system. In
1999, another soft source, RX~J0806.3+1527, was identified with a pulsation
period of 321s (Israel et al 1999, also Beuermann et al 1999), again classified
as an IP. In papers accepted within days of each other, Ramsay, Hakala \&
Cropper (2002a) and Israel et al (2002) suggested that this system was similar
to RX~J1914.4+2456, so that the 321s (5.4 min) period was the orbital period of
a double degenerate binary.

The X-ray light curves of these two systems distinguishes them from AM CVn
systems, even though the binary components may be degenerate pairs in both
cases. If the variations are taken to be the orbital period as proposed, these
are shorter than those of any known AM CVn systems. Recently Warner \& Woudt
(2002) have suggested that ES Cet (KUV01584--0939, discovered in the KISO
Schmidt UV Survey: Kondo, Noguchi \& Maehara 1984) is a short period AM CVn. At
10.3 min this is only slightly longer than that of RX~J1914.4+2456, but the
system has characteristics that are more typical of AM CVns, including strong
He emission lines in the spectrum (Wegner et al 1987), double-peaked lines
(Woudt -- private communication) and a double-humped optical light curve which
may be the result of a permanent superhump (Warner \& Woudt 2002). This implies
evidence for a disk, so ES Cet is therefore probably an AM CVn system with a
period at the short extreme of the class. This categorisation is also supported
by the fact that ES Cet was not detected in the {\it ROSAT} All Sky Survey or
PSPC pointed observations (Voges et al 2000).

In this paper we review the observed properties of RX~J1914.4+2456 and
RX~J0806.3+1527 and the models currently applied to them. We then examine how
well these explain their observed properties.

\section{Observational Characteristics}

\subsection{X-ray Spectrum}

These objects were first classified as IPs because of their short but stable
X-ray periodicities, but their spectra were completely atypical of IPs. Rather
than a hard ($>1$ keV in this context) optically thin continuum without any
soft ($<1$ keV) optically thick component, their spectra showed {\it only} a
soft optically thick component. Even `soft' IPs such as PQ Gem show a hard
component (Duck et al 1994). No hard component was evident in the {\it ROSAT}
data (Motch \& Haberl 1995). Much deeper hard X-ray observations of
RX~J1914.4+2456 were carried out by Ramsay et al (2000) with {\it ASCA}: they
measured an upper limit of the bolometric luminosity of this component of
$10^{-7}$ with respect to the soft component. Recent stringent upper limits
using {\it Chandra} have also been reported by Israel et al for RX~J0806.3+1527
(these proceedings).

\subsection{Frequency Analysis of Light Curves}

When Cropper et al (1998) Fourier analysed their {\it ROSAT} HRI data of
RX~J1914.4+2456, only one period (with its harmonics) was evident in the data,
with the value of 569.38s reported by Motch \& Haberl (1995). The same was
found in the optical by Ramsay et al (2000). A similar analysis of
RX~J0806.3+1527 X-ray data by Israel et al (1999) and optical data by Israel et
al (2002) and Ramsay et al (2002a) also found only one period with its
harmonics (but see below for further discussions). Again this was atypical of
IPs, which generally show a multiplicity of periods at combinations of the
orbital and white dwarf rotation periods.

\subsection{X-ray Modulation Profile}

The folded X-ray light curves (on the detected period) are remarkably similar
in both systems (Cropper et al 1998, Burwitz \& Reinsch 2001) -- see
Figure~1. These are approximately sawtooth, with a more rapid rise than
decline. In more detail, they show a sharp rise, followed by a slower rise to
maximum, then a slower decline, with a final smaller drop back to
minimum. There are {\it no} detected X-rays for approximately half of the
period. At different epochs the light curves of RX~J1914.4+2456 change slightly
(Ramsay et al 2000), but the sawtooth profile remains. The shape of the X-ray
modulation profile is unlike that seen in any other IP, `soft' or normal.

Cropper et al (1998) argued that the shape of the X-ray emission, and its
absence over half of the period, required a small emission region moving into
and out of view. The accretion had to be far from symmetric about the rotation
axis in order to prevent a lower region rotating into view as the upper rotated
over the limb of the star; they considered absorption could not be the cause,
and that pole-switching would not be possible on the observed short timescales
in the system. 

\begin{figure}
\plotfiddle{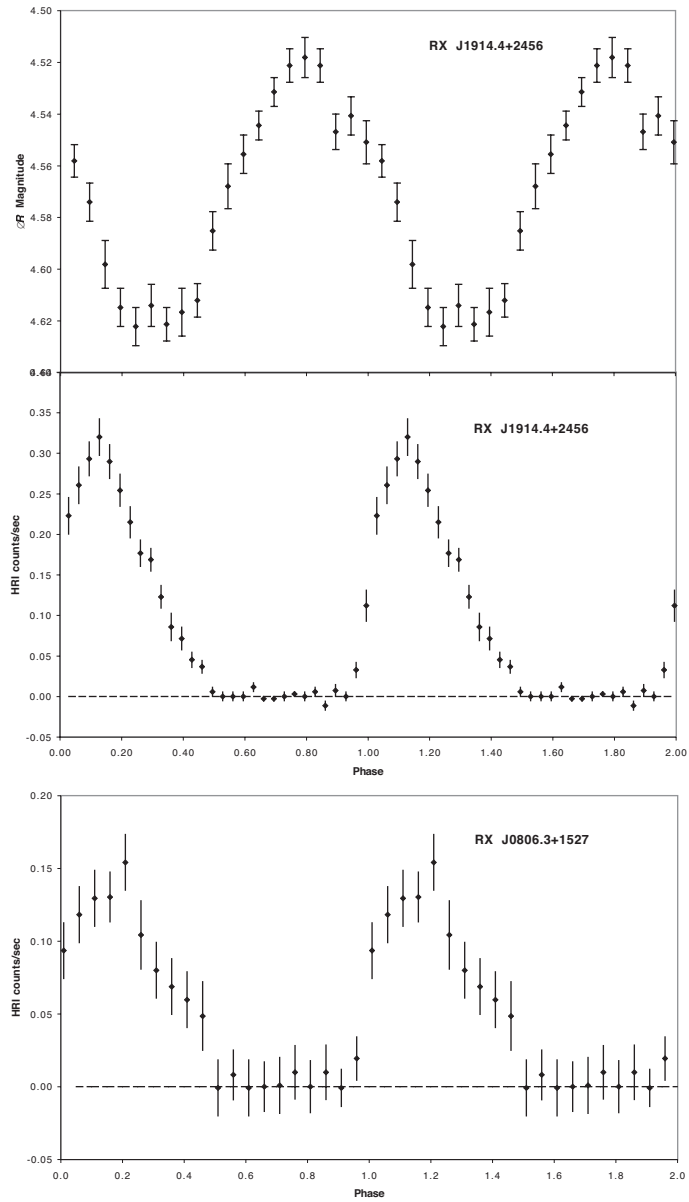}{15cm}{0}{70}{70}{-150}{0}
\caption{The {\it ROSAT} HRI X-ray light curves of RX~J1914.4+2456 (middle,
adapted from Cropper et al 1998) (1996 April) and RX~J0806.3+1527 (bottom,
adapted from Burwitz \& Reinsch 2001) (1994 October and 1995 April) phased on
the principal periodicity in each case. Also shown on the top panel is the
$R-band$ light curve for RX~J1914.4+2456 on the same phase reference from
Ramsay et al (2002b) where the magnitudes are with respect to Star 'B' of Motch
et al (1996).}
\end{figure}

\subsection{Long-term Light Curves}

Ramsay et al (2000) examined the long-term brightness of RX~J1914.4+2456 and
found that its measured brightness varied between 0.03 and 0.3 HRI cts s$^{-1}$
in the X-ray and between $I=16.6$ and $I=18.2$ in the red over a period of 6
years. Such variability in X-rays is also evident in RX~J0806.3+1527 (Israel
et al 1999) and in the optical even on short time scales (Ramsay et al 2002a,
Israel et al 2002). 

These results indicate that the emission processes in these systems are highly
variable, indicative of accretion or similar mechanism.

\subsection{Optical/IR Colours}

In order to characterise the nature of the secondary in RX~J1914.4+2456, Ramsay
et al (2000) obtained optical and infrared photometry. Their $H-K/I-K$ and
$J-H/I-K$ colour-colour diagrams located the source away from the main
sequence tracks, for all reddening values. The best description of the spectrum
is a highly reddened ($E(B-V)\sim1$) hot blackbody ($>10^{4}$K) (Ramsay et al
2002b). This reddening is consistent with the measured X-ray absorption from
cold material using {\it ASCA} and {\it ROSAT} data. If little reddening is
assumed, then a cooler spectrum is permitted, approaching 5000K. In this case,
the discrepancy between the X-ray and optical absorption requires explanation.

The situation is much clearer in the case of RX~J0806.3+1527 as this system is
much less reddened in X-rays and the optical (Israel et al 2002) and has a
spectrum characteristic of a hot blackbody ($>4\times10^{4}$K).

\subsection{Phasing of X-ray and Optical Light Curves}

Ramsay et al (2000) were able to co-phase the X-ray and optical light curves in
RX~J1914.4+2456 and found that the X-ray and optical variation was almost
anti-phased (Figure 1). The centre of the X-ray bright phase (but not the
maximum brightness) is at $\phi\sim0.25$ in their data, while the optical
maximum is at $\phi\sim0.85$ (in contrast to the X-rays, the optical/IR
variation is quasi-sinusoidal), so that the optical lags the X-rays by
$\sim0.6$. A similar value is reported in these proceedings for RX~J0806.3+1527
by Israel.

The difference in phasing between the X-ray and optical is most naturally
explained by reprocessing in a binary star model. Here the variation in the
optical emission is caused by the heating from the soft X-ray emission from
small spots on the primary (as inferred from the X-ray light curve) (see Figure
2).

\begin{figure}
\plotone{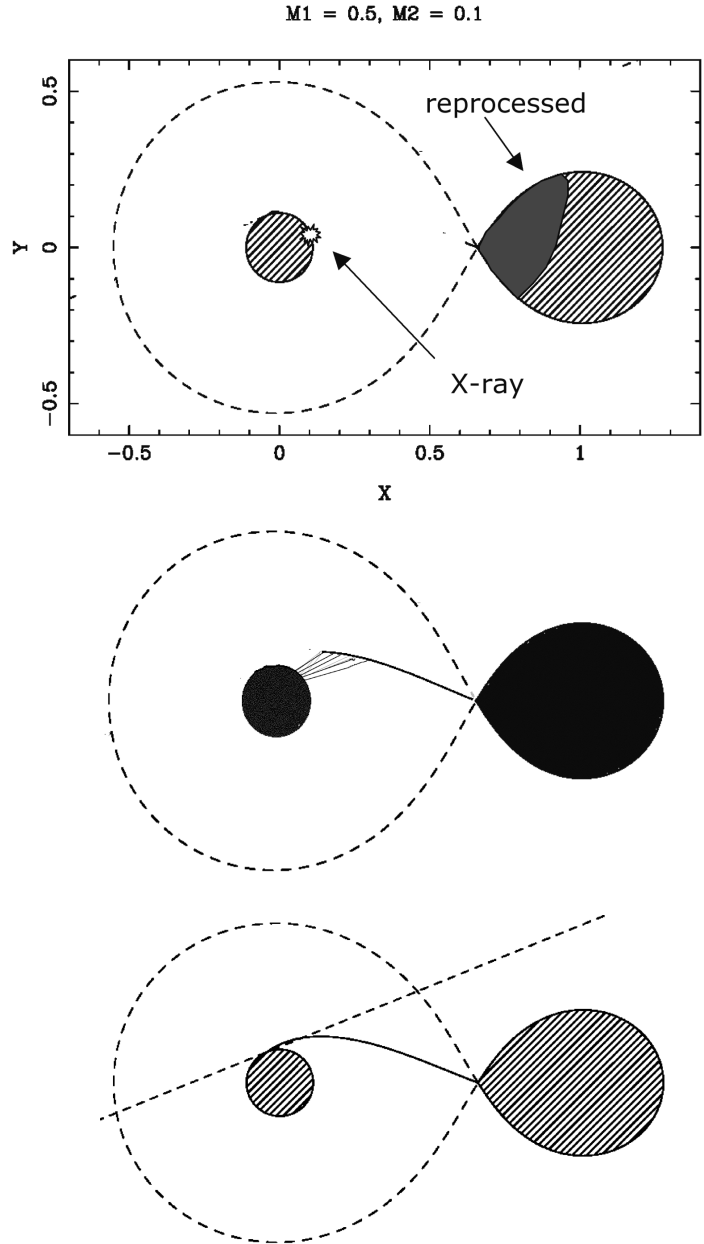}
\caption{Schematic diagrams representing the optical reprocessing by the
secondary of the X-rays from a small region on the primary (top), the
double-degenerate Polar model (centre) and the Direct Accretor model
(bottom). In the Direct Accretor, the horizon is shown from the location
of impact region, and the grazing angle of the impact is evident. All diagrams
are for a system with a period of 9.5 min as in RX~J1914.4+2456, and the
primary and secondary masses shown. Note the large size of the primary relative
to its Roche lobe. In the Electric Star model, the secondary does not fill its
Roche lobe. Diagrams adapted from Marsh \& Steeghs (2002).}
\end{figure}

\subsection{Optical Spectrum}

Despite several attempts, obtaining a spectrum of RX~J1914.4+2456 was
frustrated by the presence of nearby companions. The first optical spectrum was
presented in Ramsay et al (2002). By this time Israel et al (2002) had obtained
VLT spectra of RX~J0806.3+1527. These showed an almost featureless continuum
rising to the blue. On close inspection, the high S/N data indicated the
presence of weak $\sim5$\% emission lines, which they attributed mostly to the
Pickering series of He, with some C and Ne. The lines were broad, with
velocities of $\sim1000$ km s$^{-1}$. Norton, Haswell \&Wynn (2002) claim that
the even terms of the Pickering series lines in the spectra presented in Israel
et al (2002) are stronger, indicating the presence of {\it some H}, and
therefore ruling out a degenerate primary. This deserves further
quantification.

The spectrum of RX~J1914.4+2456 in Ramsay et al (2002) also appeared almost
featureless, with a red continuum consistent with the reddening. No emission
lines were evident in this lower S/N spectrum, but there was an absorption
feature at 5200\AA. The absence of strong emission lines in a system accreting
through Roche lobe overflow was a surprising result, and entirely atypical of
the spectrum of known IPs.\footnote{Steeghs (private communication) has
recently reported optical spectra of RX~J1914.4+2456 taken with {\it Gemini
N}. This shows absorption lines characteristic of the interstellar medium and
an overall spectrum which appears typical of a K star. The implications of
these results are being investigated by Steeghs et al.}

\subsection{Polarisation}

Ramsay et al (2000) made $I-$band polarisation measurements of RX~J1914.4+2456
and concluded that there was no circular polarisation detected on the
modulation period. The instrument used prevented a high precision measurement
of the absolute level of polarisation, but absolute levels exceeding 0.5\% and
amplitudes exceeding 0.2\% could be excluded.

\subsection{Period Changes}

Strohmayer (2002) examined the change in the 568s X-ray period of
RX~J1914.4+2456 using the archive {\it ROSAT} and {\it ASCA} data. He found
that the period was decreasing at the rate expected from gravitational losses.

\section{Models}

Currently four models have been published to describe these systems. Both
systems were originally considered to be Intermediate Polars; Cropper et al
(1998) suggested that they were double-degenerate analogs of Polars; Ramsay et
al (2002b) and Marsh \& Steeghs (2002) suggested that they might be
double-degenerate Direct Accretors and Wu et al (2002) suggested that
they might be double-degenerate systems powered by electrical current. It
should be noted that three of the four models involve a magnetic white
dwarf. We discuss these in turn below.

Other models involving neutron stars have also been considered, as the thermal
X-ray spectrum is very similar to those in the {\it ROSAT} isolated neutron
stars such as RX~J1856--37 (Burwitz et al 2002). Israel et al (1999) concluded
that such a neutron star would be only $\sim10$ pc distant, but both they and
Ramsay et al (2002a) reported negligible proper motion. Burwitz \& Reinch
(2001) argued that the absence of hard emission precluded an neutron star
interpretation for general Roche lobe overflow accretion. Ramsay et al (2002b)
also considered a neutron star model, in this case for RX~J1914.4+2456, and
concluded that this may be viable if low rates of accretion (but higher than
those expected from the accretion of interstellar material) were assumed: this
may be possible if the neutron star was accreting from material left over from
a previous phase of stellar evolution. It remains difficult to avoid the
production of a hard tail of emission, however, and pending more detailed
calculations, models in which the primary is a neutron star appear to be less
favoured.

\subsection{Intermediate Polar}

The IP interpretation is the only one of the four white dwarf primary models
with a main sequence companion. Cropper et al (1998) argued against such an
interpretation on the grounds that the X-ray light curve, and particularly the
absence of flux for half of the period and the sharp rise to maximum could not
be explained in the framework of an IP model, even if pole switching was
occurring. As noted above, further observations, for example the absence of
emission lines, the absence of a hard X-ray spectral component and the infrared
colours (the last seemingly excluding a main-sequence secondary), also militate
against an IP interpretation.

Norton et al (2002) resuscitated the IP model by investigating the
pole-switching timescales using single particle hydrodynamic (SPH) calculations
for the stream. They concluded that the rapid transitions in the X-ray light
curves could indeed be generated by pole switching in a face-on system. They
did not however address the optical/infrared colours, and the presence of
mostly He lines in the optical spectrum of RX~J0806.3+1527 was not
satisfactorily explained.

One argument against the IP model was the absence of any other periodicities in
the power spectra of these systems. These are upper limits, and other periods
may be unearthed in more sensitive future data. Indeed, despite careful
analysis, Ramsay et al (2002a) were unable to decide whether a low-level signal
in their power spectra of RX~J0806.3+1527 at 4800s was real or not. The power
spectra of Israel et al (2002) also show an enhancement at this period
(although the resolution in their plots is lower) which may mean that the
periodicity is real. In this case, asynchronous models (all except for the
first below) would be favoured, with the longer periods reflecting the orbital
period or some combination of the orbital and primary rotation periods.

Strohmeyer (2002) notes that the decrease in the 568s period in RX~J1914.4+2456
could be associated with the spinup of the primary in an IP model.

\subsection{Double-Degenerate Polar}

Because of the difficulties of the IP interpretation, and the softness of the
X-ray spectrum of RX~J1914.4+2456 and the similarity of its X-ray light curve
to those Polars accreting at one visible pole only, and the presence of only
one significant periodicity in the system, Cropper et al (1998) suggested that
the system may be the first of the double-degenerate analogs of Polars, so that
the 569s period was the orbital period of a rotationally synchronised pair of
white dwarfs (Figure 2).

Cropper et al (1998) noted that for such a compact system, the magnetic field
of the primary did not need to be as strong as that of normal Polars in order
to achieve synchronism. The predictions from the model were that there may be
detectable levels of optical circular polarisation and that the accretion flow
should be mostly He. The model successfully predicted the optical/IR colours in
the two systems (Ramsay et al 2000), and the He emission (Israel et al 2002)
but the absence of strong emission lines from an accretion stream was a
surprise. In retrospect, the small dimensions of these systems renders the
stream to be small, and this and the hotter white dwarf secondary could cause
the line emission from the stream to be feeble by comparison with normal
Polars.

The absence of circular polarisation was also seen as a possible difficulty for
the double-degenerate Polar model by Ramsay et al (2002b). However they point
out that magnetic fields less than a few MGauss would not generate 
the high levels of polarisation seen in usual Polars.

While the ratio of soft X-rays to hard in some Polars can be high (see Ramsay
\& Cropper, these proceedings), in general some optically thin hard component
from the post-shock region is detected. The absence of any detection of a hard
component (Ramsay et al 2000, Israel et al, these proceedings) therefore poses
some difficulties. One possibility is that the flow between the secondary and
primary is subsonic all of the way (a `settling solution'). If this is the case
the temperature structure within the stream should be evident as it appears
over the limb of the star; moreover, in this case it is difficult to imagine
how X-ray emission is absent for half of the orbital period, even if most of
the emission is near the base of the accretion flow onto the white dwarf.

\subsection{Double-Degenerate Direct Accretor}

The double-degenerate Direct Accretor (sometimes loosely called
double-degenerate Algol) model was proposed independently by Marsh \& Steeghs
(2002) and Ramsay et al (2002b). In this case, neither of the white dwarfs is
magnetic, and mass is lost from the secondary by Roche Lobe overflow in a
ballistic stream. The larger relative size of the primary and its Roche Lobe in
these compact systems is such, however, that the stream impacts directly onto
its surface, ahead of the line of centres (Figure 2). For smaller (more
massive) primaries, the stream may miss the surface and intersect with iself,
circularising and forming an accretion disk as in the standard AM CVn systems.

The above papers used slightly different constraints for determining the
allowed parameter space for such a model to be viable, but both agreed in their
analysis that the primary mass could not exceed $\sim0.55$ M$_{\odot}$. Marsh
\& Steeghs (2002) took the point of view that this was consistent with known
primary masses in interacting binaries, while Ramsay et al (2002b) considered
the restricted parameter space to be a concern, searching as they were for
other models which avoided the restricted magnetic field in the
double-degenerate Polar model.

It should be noted that unless the primary is substantially less massive than
0.55 M$_{\odot}$, the stream impacts on the far side of the primary from the
secondary, and the impact is at an oblique angle. In this case, the secondary
is below the horizon, and it cannot be the source of the optical modulation
through reprocessing of the X-ray flux. In this case Marsh \& Steeghs (2002)
suggested that the optical modulation results from the cooling material
downstream of the X-ray heated region (the non-magnetic primary would not be
synchronised). The implications of this are that there is no fixed relationship
between the X-ray and optical phasing: this would differ for different
accretion rates; it would be different from system to system, depending on
the rotation rate of the primary; there should be a systematic advancement in
the phasing of the optical/IR light curves towards longer wavelengths, and
because of the prograde rotation of the primary, the X-ray light curve should
have a slow rise and steep decline, together with a strong X-ray spectral
signature. On most of these points such an explanation can be excluded, so that
this model requires a very low mass primary.

\subsection{Electric Star}

In response to the difficulties with all of the models above, Wu et al (2002)
suggested that RX~J1914.4+2456 was powered by electrical current. They noted
that a white dwarf secondary star moving in the magnetic field of the primary
would develop an electrical potential across its diameter. Should there be any
ionised material between the two stars, an electrical circuit would be set up
each side of the orbital plane: down the magnetic field lines from one side of
the secondary to the primary and back down field lines to the other side. The
major source of resistance in this circuit is in two small regions where the
magnetic field lines intersect the surface of the primary near the magnetic
poles. This is where almost all of the energy is dissipated.

This unipolar inductor system was based on that observed directly in the
Jupiter-Io interaction (Clarke et al 1996) and proposed white dwarf-planet
interactions (Li, Ferrari \& Wickramasinghe 1998) but scaled up to take account
of the much shorter orbital period, potentially much larger primary magnetic
field and relatively much larger secondary in the double-degenerates. Wu et al
(2002) found that the total energy dissipated at the spots in the white dwarf
primary atmosphere exceeded $10^{32}$ erg s$^{-1}$ for a fractional
asynchronism of 0.001 between the rotation of the primary and that of the
orbit. The energy reservoir in the system is the revolving and rotating binary
pair; gravitational radiation robs the orbit of energy, so that the orbital
separation shrinks and the period shortens, but does not act
significantly on the stellar rotation. The primary is therefore being
contnually spun up through the Lorentz torques caused by the electrical
currents, and from the point of view of the secondary it moves slowly
retrograde. However, the footpoints of the field lines remain locked in the
frame of the binary as the primary rotates, as they are the lines which thread
the secondary.

The secondary star in the electric star model does not fill its Roche lobe, and
there is no accretion. The model therefore naturally explains the absence of
strong emission lines as may be expected from a stream: any lines that are
evident will be produced by the heated secondary, and ought therefore to be
rotationally broadend. The magnetic field required on the primary is moderate,
$\sim 1$ MG, and there is no shock, so the levels of optical circular
polarisation should be small and there should be no hard X-rays produced. There
are natural instabilities in the system caused by induced magnetic fields (from
the large curents flowing in the circuit) that modify the field of the primary:
these could be the origin of the long- and short-term variability seen in the
systems. Such detatched white dwarf pairs in which one of the stars has a
moderate magnetic field will be natural end-points of many binary systems, and
the extreme mass ratios required for Roche Lobe overflow are not required in
this model. Furthermore, the steep rise and more shallow decline in X-ray
brightness is in the right sense as (unlike the Direct Accretor) the
primary is rotating retrograde.

Strohmeyer (2002) notes that the binary period will decrease in the absence of
conservative mass transfer (Wu et al 2002), but increase if conserved mass
transfer takes place (Rappaport, Joss \& Webbink 1982). The observed decrease
in period consistent with gravitational radiation losses therefore favours a
non-accretor interpretation and hence the Electric Star model.

The Electric Star model appears to be consistent with the observational facts
for RX~J1914.4+2456 and RX~J0806.3+1527 but Wu et al (2002) note that the
lifetime of this phase would be short at $\sim 10^{3}$ years. Marsh \& Steeghs
(2002) considered this a serious difficulty given the space density that can be
inferred from two systems. However, this difficulty may be overcome, as the
system may be oscillating between a unipolar inductor phase and an accreting
phase many times through the lifetime of the binary, which may last
$10^{6}-10^{7}$ years.

\subsection{Comparison}

We summarise in Table~1 the degree with which the different models are
consistent with the observational characteristics of the two systems.

\begin{table}
\begin{tabular}{|l|cccc|} \hline
   & Intermediate & DD & DD & Electric \\
   & Polar   & Polar & Direct & Star   \\
   &    &    & Accretor &   \\ \hline
Absence of hard X-rays     & $\times$ & $\sim$  & $\surd$ & $\surd$ \\
Only one modulation period & $\times$ & $\surd$ & $\surd$ & $\surd$ \\
Optical/IR colours         & $\times$ & $\surd$ & $\surd$ & $\surd$ \\
Phasing of X-ray and optical & $\sim$ & $\surd$ & $\sim$  & $\surd$ \\
Shape of X-ray modulation  & $\surd$  & $\surd$ & $\sim$  & $\surd$ \\
Absence of Polarisation     & $\surd$  & $\sim$  & $\surd$ & $\surd$ \\
Absence of H               & $\sim$   & $\surd$ & $\surd$ & $\surd$ \\
Absence of strong emission lines & $\times$ & $\sim$ & $\sim$ & $\surd$ \\
Long-term variability      & $\surd$  & $\surd$ & $\surd$ & $\surd$ \\
Period change              & $\surd$  & $\times$ & $\times$ & $\surd$ \\
Lifetime                   & $\surd$  & $\surd$  & $\surd$ & $\sim$ \\
\hline
\end{tabular}
\caption{Model comparison for the four models containing a white dwarf
primary. $\times$, $\sim$ and $\surd$ symbols indicate increasing degrees of
compliance with the observational data.}
\end{table}

Of the four models, the IP interpretation has the most difficulties. The
absence of strong emission lines, the optical/IR colours (particularly in 
RX~J0806.3+1527 where the uncertainties introduced by the reddening are not
important) and the absence of a hard optically thin X-ray component all apppear
to be insurmountable difficulties, even if the claim by Norton et al (2002)
that the shape of the X-ray light curve can be caused by pole switching is
accepted. 

The double degenerate Polar and Direct Accretor models provide a reasonably
good description of the observed data. Both are inconsistent with the period
change in RX~J1914.4+2456 measured by Strohmayer (2002), who points out that
although there may be other factors affecting the period, those identified in
normal CVs are generally absent when the secondary is degenerate. Period
changes will become increasingly evident as time passes, especially in the case
of RX~J0806.3+1527, and this promises to be a powerful way to discriminate
between the accreting and Electric Star models.

Both the double degenerate Polar and Direct Accretor models ought to have some
signature of the accretion stream at a low level. It remains to be seen whether
the lines seen by Israel et al (2002) in RX~J0806.3+1527 can be explained fully
by emission from the heated face of the secondary, or whether they require an
additional component from the stream. This can be solved using Doppler
Tomography.

The Direct Accretor model suffers from the requirement that the primary star
needs to be relatively low mass, and, if a reprocessing model is accepted for
the optical component, the primary mass is uncomfortably low in order for the
secondary to be illuminated by the X-ray emitting regions. The suggestion by
Marsh \& Steeghs (2002) that the optical is the cooling tail of the X-ray
emitting region on the primary is not particularly satisfactory on the grounds
of the X-ray modulation profile, the absence of phase shifts between the
different optical/IR bands and the constant phase difference at different
accretion rates; also the similar X-ray/optical phasing observed in the two
systems (Ramsay et al 2000, Israel, these proceedings). On the other hand, the
absence of polarisation in the double degenerate Polar model may not be too
significant, given the modest field strengths required to achieve synchronism.

Of all the currently envisaged scenarios, the Electric Star model matches the
observations most closely. The main disadvantage of the model is that the
lifetime is predicted to be too short given the observed space
density. However, the calclation in Wu et al (2002) used magnetic field
strengths and departures from asynchronism which aimed primarily to demonstrate
that sufficient power was available from these systems, rather than using
parameters which maximised the lifetime given an observed luminosity. How much
the lifetime is a difficulty also depends on how many close white
dwarf pairs are expected from evolutionary calculations, which at present are
not sufficiently constraining. Further open issues of the model are that it
relies on an initial disturbance into asynchronism, and that the effects of the
induced currents on the magnetic field of the primary still need to be
calculated. Perhaps the main difficulty faced with the model is one of
unfamiliarity in the context of accreting binaries, thus requiring an extra
burden of proof.

\section{Gravitational Radiation}

Ramsay et al (2000) made the point that the accretion in these systems was
driven by gravitational radiation. Wu et al (2002) estimated that the power of
the gravitational radiation in such systems can easily reach $\sim10^{36}$ erg
s$^{-1}$. A number of authors (for example Strohmeyer 2002) have pointed out
that they will radiate at frequencies $2/P_{orb}$ at the centre of the band at
which the proposed {\it LISA} mission will be
sensitive. Because of the large gravitational radiation power, the large strain
amplitudes ($\sim10^{-21}$) they generate in this band and their proximity will
cause these systems to be the brightest presently known constant sources in the
gravitational wave sky, and as such they will be among the first to be
observed. Moreover, these will be sources with relatively well-known source
properties (masses, orbital periods) and will therefore be fundamentally
important in early gravitational wave studies.

\section{Conclusions}

With the discovery of RX~J0806.3+1527, so similar to RX~J1914.4+2456, it now
appears more clear that the secondary in these systems is a degenerate dwarf,
and that the main period in these systems is the orbital period. The
uncertainties on the nature of the secondary introduced by the reddening
correction to the optical/IR colours in RX~J1914.4+2456 are negligible in
RX~J0806.3+1527, and the shorter period makes it impossible to squeeze in a
main sequence star. The similar phasing between the X-ray and optical in the
two systems and its stability from epoch to epoch suggests that the source of
the optical variation is reprocessing from the secondary. This again argues for
a binary model.

A neutron star primary is not entirely excluded as a possibility, but because
of the gravitational bending of the emission into fan-shaped beams, the sharp
X-ray pulse profiles are difficult to generate if these are assumed to arise in
hot polar caps. The systems need to be extremely close in order to attain the
measured X-ray intensities from small regions on a neutron star: the reddening
in RX~J1914.4+2456 and the lack of proper motion of RX~J0806.3+1527 (Ramsay et
al 2002a, Israel et al 2002) makes this unlikely. The absence of a hard
component rules out accretion onto a neutron star, except possibly for a very
narrow parameter regime.

For these reasons models with a white dwarf primary are preferred. Of the four
models considered so far, it would be unwise to exclude any except the IP
interpretation. Of these, the Electric Star model provides the best agreement
with the observed characteristics of these systems, but their short lifetimes
remain an issue given the space density that can be inferred from two systems.

Looking forward, progress in understanding these systems will depend on more
being found. All of the double-degenerate models are viable and together with
the disk-accreting AM CVn systems, it may be that larger samples indicate that
there is a melange of double degenerate systems with ultrashort binary
periods. It may even be possible for individual binaries to move from Electric
Star to accreting systems, as alluded to above, or from double-degenerate
Polars to double-degenerate IPs. One of the main discriminants between the
models is the presence and extent of an accretion stream, and Doppler
tomography will be an important tool in identifying if a stream contributes to
the line emission. A further important tool, as pointed out by Strohmayer
(2002) is the investigation of the period changes in these systems; more
constraints on their space density from evolutionary models will also prove
valuable.

\section{Acknowledgements}

Mark Cropper is grateful to the Royal Society for travel support.


\begin{references}
\reference Beuermann, K., Thomas, H.-C., Reinsch, K., Schwope, A. D., 
Tr\"{u}mper, J, Voges, W., 1999, A\&A, 347, 47
\reference Burwitz. V., Reinsch, K. 2001, {\it X-ray astronomy : stellar
endpoints, AGN, and the diffuse X-ray background}, Bologna, Italy, eds White,
N. E., Malaguti, G., Palumbo, G., AIP conference proceedings, 599, 522
\reference Burwitz, V., Haberl, F., Neuhaeuser, R., Predehl, P., Tr\"{u}mper, 
J., Zavlin, V. E. 2002, A\&A, in press, astro-ph/0211536
\reference Clarke, J. T. et al, 1996, Science, 274, 404
\reference Cropper, M., Harrop-Allin, M. K., Mason, K. O., Mittaz, J. P. D.,
Potter, S. B., Ramsay, G., 1998, MNRAS, 293, L57
\reference Duck, S. R., Rosen, S. R., Ponman, T. J., Norton, A. J., Watson,
M. G., Mason, K. O. 1994, MNRAS, 271, 372
\reference Israel G.-L., Panzera M.R., Campana S., Lazzati D., Covino S.,
Tagliaferri G. 1999, A\&A, 349, L1
\reference Israel, G.-L., Hummel, W., Covino, S., Campana, S., Appenzeller, I.,
G\"{a}ssler, W., Mantel, K.-H., Marconi, G., Mauche, C. W., Munari, U., 
Negueruela, I., Nicklas, H., Rupprecht, G., Smart, R. L., Stahl, O., 
Stella, L. 2002, A\&A, 386, L13
\reference Kondo, M., Noguchi, T., Maehara, H. 1984, {\it Ann. Tokyo
Astron. Obs.}, 20, 130
\reference Li, J., Ferrario, L., Wickramasinghe, D. T., 1998, ApJ, 503, L151
\reference Marsh, T., Steeghs, D. 2002, MNRAS, 331, L7
\reference Motch, C., Haberl, F. 1995, {\it Cape Workshop on Magnetic
Cataclysmic Variables}, ASP Conf. Ser. 85, eds Buckley, D. A. H., Warner, B., p
109
\reference Motch, C., Haberl, F., Guillout, P., Pakull, M., Reinsch, K.,
Krautter, J. 1996, A\&A, 307, 459
\reference Norton, A. J., Haswell, C. A., Wynn, G. A. 2002, astro-ph/0206013
\reference Ramsay, G., Cropper, M., Wu, K., Mason, K. O., Hakala, P. 2000,
MNRAS, 311, 75
\reference Ramsay, G., Hakala, P., Cropper, M. 2002a, MNRAS, 332, L7
\reference Ramsay, G., Wu, K., Cropper, M., Schmidt, G., Sekiguchi, K.,
Iwamuro, F., Maihara, T., 2002b, MNRAS, 333, 575
\reference Rappaport, S., Joss, P. C., Webbink, R. F. 1982, ApJ, 254, 616
\reference Smak, J., 1967, Acta Astron., 17, 255
\reference Strohmayer, T., 2002, ApJ, 581, 577
\reference Voges, W., Aschenbach, B., Boller, Th., Brauninger, H., Briel, U., 
Burkert, W., Dennerl, K., Englhauser, J., Gruber, R., Haberl, F., Hartner, G.,
Hasinger, G., Pfeffermann, E., Pietsch, W., Predehl, P., Schmitt, J., 
Tr\'{u}mper, J., Zimmermann, U. 2000, VizieR On-line Data Catalog: IX/29.  
Originally published in: Max-Planck-Institut fur extraterrestrische Physik, 
Garching (2000)
\reference Warner, B., 1995, Ap\&SS, 225, 249
\reference Warner, B., Woudt, P. A. 2002, PASP, 792, 129
\reference Wegner, G., McMahon, R. K., Boley, F. I., 1987, AJ, 94, 1271
\reference Wu, K., Cropper, M., Ramsay, G., Sekiguchi, K. 2002, MNRAS, 331, 221
\end{references}
\end{document}